\documentclass[prb,twocolumn,superscriptaddress,showpacs,floatfix,amsfonts]{revtex4}
\usepackage{graphicx,graphics,color,epsfig}
\usepackage{bm}
\usepackage{amsmath}
\usepackage{amssymb}
\usepackage{epstopdf}
\begin{document}
\preprint{}
\title{Spectral Properties of $\delta$-Plutonium: Sensitivity to $5f$ Occupancy}
\author{Jian-Xin Zhu}
\affiliation{Los Alamos National Laboratory,
Los Alamos, New Mexico 87545, USA}
\author{A. K. McMahan}
\affiliation{Lawrence Livermore National Laboratory, Livermore, California 94550, USA }
\author{M. D. Jones}
\affiliation{University at Buffalo, SUNY, Buffalo, New York 14260, USA }
\author{T. Durakiewicz}
\affiliation{Los Alamos National Laboratory,
Los Alamos, New Mexico 87545, USA}
\author{J. J. Joyce}
\affiliation{Los Alamos National Laboratory,
Los Alamos, New Mexico 87545, USA}
\author{J. M. Wills}
\affiliation{Los Alamos National Laboratory,
Los Alamos, New Mexico 87545, USA}
\author{R. C. Albers}
\affiliation{Los Alamos National Laboratory,
Los Alamos, New Mexico 87545, USA}

\date{\today}

\begin{abstract}

By  combining the local density approximation (LDA) with dynamical
mean field theory (DMFT), we report a systematic analysis of
the spectral properties of $\delta$-plutonium with varying
$5f$ occupancy. 
The LDA Hamiltonian is extracted from a tight-binding (TB) fit to
full-potential linearized augmented plane-wave (FP-LAPW) calculations.
The DMFT equations are solved by the exact quantum Monte Carlo (QMC) method
and by the Hubbard-I approximation.  
We demonstrate
strong sensitivity of the spectral
properties to the $5f$ occupancy, which suggests using this
occupancy as a fitting parameter in addition to the Hubbard $U$.
By comparing with 
photoemission data, we conclude that the 
``open shell'' $5f^{5}$ configuration gives the best agreement, 
resolving the controversy over $5f$ ``open shell'' versus 
``close shell'' atomic configurations in $\delta$-Pu.

\end{abstract}
\pacs{71.20.Gj,	
71.27.+a, 
75.20.Hr,  
79.60.-i	
}
\maketitle

\section{Introduction}

From a consideration of its condensed matter physics properties,
crystal structure, and metallurgy, plutonium is probably
the most complicated element in the periodic
table,~\cite{OJWick67,DAYoung91,NGCooper00,RCAlbers01,SSHecker04,RCAlbers07}
including a phase diagram with six allotropic phases. 
The low-temperature monoclinic $\alpha$-phase is stable up to
395 K while the face-center-cubic (fcc) $\delta$-phase
is stable between 592 and 724 K. 
Furthermore, in stabilized alloys the $\alpha$ to $\delta$ phase 
transformation of Pu has a significant volume expansion, 
with the $\delta$-phase 25\% larger in volume than the  $\alpha$-phase. 
This behavior is related to the special position of Pu in the 
periodic table, which is at the boundary between the light 
actinides that have itinerant  5$f$ electrons and the heavy 
actinides with localized 5$f$ electrons.  
In this situation the electrons are in a very strongly
correlated state where even the best conventional LDA 
band-structure calculations cannot predict its unique properties, 
e.g., the $\delta$-phase volume~\cite{PSoderlind94,MDJones00}
in the experimentally observed nonmagnetic state. 
This failure has stimulated numerous attempts to include
additonal electronic correlation effects.
Although LDA theory has repeatedly
claimed~\cite{IVSolovyev91,YWang00,PSoderlind01,PSoderlind02,PSoderlind04} 
that the thermal expansion in Pu is a consequence of magnetism,
this is in striking contradiction with experimental data.~\cite{JCLashley05}  
Several research groups~\cite{JBouchet00,DLPrice00,SYSavrasov00}
have applied the LDA+$U$ method to include more $f$-$f$ correlation 
energy. By adjusting the on-site Coulomb repulsion energy ($U$) 
appropriately, it has been possible to fit to the experimental 
$\delta$-phase volume.  
The LDA+$U$ calculations also indicated an instability of 
$\delta$-Pu toward an antiferromagnetic ground state.  
More recent calculations based on the ``around-mean-field'' 
LSDA+$U$ including the spin-orbit interaction~\cite{ABShick05} 
or the fully-localized-limit LSDA+$U$ quenched the spin 
polarization through an $\approx f^{6}$ configuration, 
and obtained the correct $\delta$-phase 
volume due to a weak exchange interaction.~\cite{AOShorikov05} 
To predict the correct volume, a mixed-level model has  also been 
proposed by one of us together with co-workers.~\cite{JMWills04}  
In this model,  four of the five 5{\em f} electrons in the 
$\delta$-phase were constrained to be localized  and were not 
allowed to hop from site to site, but could hybridize with the 
conduction electrons.  A major advance came when a new 
calculation scheme was proposed to merge LDA-based methods  
with DMFT.~\cite{VIAnisimov97,GKotliar06} 
DMFT~\cite{ThPruschke95,AGeorges96}  is a many-body technique that is able to treat the 
band- and atomic-like aspect simultaneously when applied to plutonium.  
Within the LDA+DMFT approach,~\cite{GKotliar06,SYSavrasov01}  
the origin of substantial volume expansion was explained in terms
of the competition between Coulomb repulsion and kinetic energy. 
Thus, two dramatically different pictures for the non-magnetism
of $\delta$-Pu have emerged. 
In the ``5$f^{6}$'' 
picture,~\cite{ABShick05,AOShorikov05,LVPourovskii05,AShick06}
by including the spin-orbit interaction, one starts with a 
closed $f^{5/2}$ atomic subshell fully filled 
with six electrons while the $f^{7/2}$ subshell is empty, 
making Pu magnetically inert. 
In the  ``5$f^{5}$'' picture,~\cite{SYSavrasov01,JHShim06} 
one starts with an open $f^{5/2}$ atomic subshell filled with five
electrons, resulting in a magnetic moment that should be screened
by the $sd$ valence electrons.
Therefore, more studies are necessary to determine 
which picture will prevail.  

In this paper, we report
a systematic LDA+DMFT study of the spectral 
properties of $\delta$-Pu with varying $5f$ occupancy. 
Throughout this work, the LDA part of Hamiltonian is
determined from a new  TB fit to FP-LAPW calculations.  
The DMFT equations are solved using quantum Monte Carlo (QMC) 
simulations as well as the more approximate Hubbard-I method. 
These provide an accurate characterization of spectral properties of $\delta$-Pu. 
It is found that the $5f$ spectral density of Pu is very sensitive to its occupancy. 
Good agreement is found with photoemission spectroscopy (PES) measurements
when about five electrons occupy the $f^{5/2}$ subshell, 
in support of the second picture.  
Other recent work appears also to be in support of this picture.  
Measurements of the branching ratio in $5d$ to $5f$
  transitions favor a $5f$ count closer to $5$ than
  $6$,~\cite{Moore06,vanderLaan04} while subsequent LDA+DMFT
  calculations~\cite{JHShim06} using a vertex-corrected one-crossing
  approximation for the auxilliary impurity problem yield results
  for the branching ratio in agreement with experiment~\cite{Moore06}
  and suggest a $5f$ occupancy of $5.2$.
Similarly, specific heat calculations 
carried out using an LDA+DMFT method with a perturbative $T$-matrix 
and fluctuating exchange approach to the auxiliary impurity 
problem also favor the $\sim 5f^5$ configuration.~\cite{LVPourovskii07}

The remainder of this paper is organized as follows: Theoretical 
methods are given in Sec.~\ref{SEC:Model}; numerical results for the 
$\delta$-Pu are presented in Sec.~\ref{SEC:Spectrum}; and then a 
summary follows in Sec.~\ref{SEC:Summary}

\section{Theoretical methods}
\label{SEC:Model}

Most LDA+DMFT methods to date have been implemented in a
basis of linear muffin-tin orbitals (LMTO) within the atomic sphere approximation
(ASA), or in its full potential
version. 
Here we use an alternative version of the LDA+DMFT method, which is based on a recently developed TB theory.~\cite{MDJones02} 
In this representation, the full second-quantized Hamiltonian is written as:
\begin{eqnarray}
\hat{H}& = & \sum_{\mathbf{k},ljm_{j},l^{\prime}j^{\prime}m_{j}^{\prime}}[H^{0}(\mathbf{k})]_{ljm_{j},l^{\prime}j^{\prime}m_{j}^{\prime}}\hat{c}^{\dagger}_{\mathbf{k}ljm_{j}}\hat{c}_{\mathbf{k}l^{\prime}j^{\prime}m_{j}^{\prime}} \nonumber \\
&& + \frac{U_{f} }{2}\sum_{i,jm_{j}\neq j^{\prime}m_{j}^{\prime}}\hat{n}_{ifjm_{j}}\hat{n}_{ifj^{\prime}m_{j}^{\prime}}\;.
\label{EQ:Hamil}
\end{eqnarray}
Here $\mathbf{k}$ are the Brillouin-zone wavevectors; $i$ are lattice site
indices; $l$ is the orbital angular momentum; $j$ is the total angular
momentum; $m_{j}=-j,-j+1,...,j-1,j$; $\hat{n}_{ifjm_{j}}\equiv \hat{c}^{\dagger}_{ifjm_{j}}\hat{c}_{ifjm_{j}}$. The relevant orbitals for Pu are $7s$, $6p$, $6d$, and $5f$, and so the matrices  $\hat{H}^{0}(\mathbf{k})$ are $32\times 32$. 
They are given by
\begin{equation}
\hat{H}^{0}(\mathbf{k}) =
\hat{H}^{\text{LDA}}(\mathbf{k}) +
(\varepsilon_f-\varepsilon_f^{\text{LDA}}) \hat{I}_f \, .
\label{EQ:H0}
\end{equation}
Here the matrix $\hat{I}_f$ is zero except for $1$'s along the $14$
$f$-$f$ diagonals, and
\begin{equation}
\varepsilon_f^{\text{LDA}} = \frac{1}{14N} \sum_{\mathbf{k}}
\text{Tr}[\hat{H}^{\text{LDA}}(\mathbf{k}) \hat{I}_f ] \, ,
\label{EQ:efLDA}
\end{equation}
where $N$ is the number of $\mathbf{k}$ points in the Brillouin zone.
The matrices $\hat{H}^{\text{LDA}}(\mathbf{k})$ are orthogonalized,
single-electron Hamiltonian matrices obtained from TB fits to the
FP-LAPW calculations.~\cite{MDJones02}  In this study, strict orthogonality
is maintained between the TB orbitals (hence no overlap matrix need be included).
All matrices  are calculated at the experimental
$\delta$-Pu volume.

The Slater-Koster tables for the $sp^3d^5$ matrix elements can be
found in standard references,~\cite{JCSlater54,WAHarrison80} and we have used
an extended formalism  for a unified treatment
including additional matrix elements involving $f$-electrons.~\cite{McMahan98}
Typical TB calculations are then reduced to using TB as an interpolation
scheme; the matrix elements are
determined by fitting to {\it ab-initio} calculated quantities such as the
total energy and band energies.
In this study TB inter-site parameter values are evaluated at inter-atomic distances out to the 
fifth nearest neighbor, resulting in 100 inter-site parameters and 4 on-site parameters.
The effect of the spin-orbit interaction is included as a perturbation, with its parameters kept fixed across the bands, 
although evaluated at the most important energies, namely at the respective centers of gravity of the 
occupied state density for each orbital type.~\cite{MDJones07} 
The spin-orbit coupling adds 3 more parameters to the TB fit, resulting in a
total of 107 parameters (parameter values are available on request from the
authors). The quality of the present TB fit for $\delta$-Pu is comparable to that shown 
elsewhere for fcc U.~\cite{MDJones02}


As customary in LDA+DMFT calculations,
Eqs.~(\ref{EQ:Hamil}-\ref{EQ:efLDA}) presume that the
LDA results are sufficient for both the off-diagonal elements of
$\hat{H}^{0}(\mathbf{k})$ (hybridization) and the diagonal elements
of the assumed uncorrelated $spd$ electrons.  For the correlated
$f$ electrons one needs of course the Coulomb interaction as
seen in Eq.~(\ref{EQ:Hamil}), but in addition, since the LDA
site energy defined by Eq.~(\ref{EQ:efLDA}) includes Coulomb
contributions, it must be replaced by the true or bare site energy
$\varepsilon_f$ as in Eq.~(\ref{EQ:H0}).  Both $\varepsilon_f$ and
$U_f$ may be estimated from the dependence of the LDA site energy
$\varepsilon_f^{\text{LDA}}(n_f)$ on $f$ occupation $n_f$ via LDA
constrained occupation calculations,\cite{McMahan88,Anisimov91}
and comparison to spectroscopic data for ions suggests uncertainties
of $\pm 1$ eV in such results for both quantities.\cite{McMahan88}
Typical values of $U_f$ are around 4 eV for $\delta$-Pu, as we
also find, and this value is assumed throughout the present paper.
Given the sensitivity of $n_f$ to $\varepsilon_f$, on the other hand,
we have chosen to adjust the parameter $\varepsilon_f$ in the present
work as a natural and convenient means of exploring the sensitivity
of the $\delta$-Pu spectra to $5f$ occupancy.~\cite{ASAresults}
This adjustment
amounts to varying the double counting term in conventional LDA+$U$ or LDA+DMFT
terminology.

Within the DMFT, the lattice problem Eq.~(\ref{EQ:Hamil}) is mapped onto a 
multi-orbital quantum single impurity problem subject to a self-consistency condition:
\begin{equation} 
\hat{\mathcal{G}}^{-1}(i\omega_{n}) = \hat{G}_{\text{loc}}^{-1}(i\omega_{n}) + \hat{\Sigma}(i\omega_{n})\;.
\end{equation}
Here $\hat{\mathcal{G}}(i\omega_{n})$ is the
Weiss function, $\hat{\Sigma}(i\omega_{n})$ is a
$\mathbf{k}$-independent self-energy, and the local Green's
function is defined as $\hat{G}_{\text{loc}}(i\omega_{n})
=\sum_{\mathbf{k}}\hat{G}_{\mathbf{k}}(i\omega_{n})/N$, where
the lattice Green's function reads
\begin{equation}
\hat{G}_{\mathbf{k}}(i\omega_{n}) = [(i\omega_{n} + \mu)\hat{I} - \hat{H}^{0}(\mathbf{k}) -
\hat{\Sigma}(i\omega_{n})]^{-1}\;,
\end{equation}
with  $\hat{I}$ the $32\times32$ unit matrix and $\mu$ the chemical
potential. Since we have restricted strong correlation to the $f$
orbitals only, it is customary to take $\hat{\Sigma}$ to have nonzero
elements only within the $14\times14$ $f$-$f$ block, which permits
a similar reduction in the quantum  single-impurity problem itself.
Furthermore, we neglect the off-diagonal elements in the self-energy
and,  by ignoring the crystal-field effect in the impurity problem,
treat the $5f_{j=5/2}$ levels as one 6-fold degenerate level and the
$5f_{j=7/2}$ as another 8-fold degenerate level. The only nonzero
$f$-$f$ block self-energy reduces to
\begin{equation} 
 \Sigma^{ff}_{jm_{j},j^{\prime}m_{j}^{\prime}}(i\omega_{n})=\Sigma^{ff}_{j}(i\omega_{n}) \delta_{jj^{\prime}}\delta_{m_{j}m_{j}^{\prime}}\;,
 \end{equation}
and we need only find functions for the two spin-orbit states $j=5/2$
and $j=7/2$. Note that when no spin-orbit coupling is included,
$\Sigma^{ff}_{5/2}(i\omega_{n})=\Sigma^{ff}_{7/2}(i\omega_{n})$.
Correspondingly, the local Green's functions is dictated by:
\begin{equation} 
G^{ff}_{\text{loc},jm_{j},j^{\prime}m_{j}^{\prime}}(i\omega_{n})=\frac{\delta_{jj^{\prime}}\delta_{m_{j}m_{j}^{\prime}}}{(2j+1)N} \sum_{\mathbf{k}}\text{Tr}_{fj}\{\hat{G}_{\mathbf{k}}(i\omega_{n})\}\;.
\end{equation}
It should also be noted that we neglect the crystal-field
effects only at the auxiliary impurity problem, all such effects
are still retained in $\hat{H}^{0}(\mathbf{k})$,
where hybridization dominates anyway.

To solve the auxiliary impurity problem, we have employed both
the Hubbard-I approximation and an implementation of the QMC
method that has been used to study properties of the compressed
lanthanides.~\cite{KHeld01,AKMcMahan03,AKMcMahan05}  The former
method is more approximate but much faster while the latter is
more rigorous but computationally more expensive.  The Hubbard-I
approximation is also useful in various ways:  it is more applicable in
the large-volume atomic limit, where it gives insight into the
atomic aspects of Pu; it provides
a good initial guess for the DMFT(QMC) simulations; and it makes it possible
to examine the full $f$-$f$ Coulomb interaction
(all four Slater integrals $F^k$) and its consequent term structure
in DMFT(HI) in contrast to DMFT(QMC), where it is still
difficult to go beyond just the standard Hubbard $U_f\equiv F^0$.
Unless specifically noted otherwise, all DMFT(HI) and DMFT(QMC)
results reported in this paper include only $U_f$ in addition
to the spin-orbit intraction.  For the DMFT(QMC) calculations,
we used 6000 sweeps per QMC iteration for the imaginary time segments
$L=112$ at a temperature $T=632\;\text{K}$, which is about as low
a temperature as is practical at the present time with QMC, and
performed at least 100 iterations.  To improve the input data needed for the
maximum entropy (MaxEnt) analytical continuation,~\cite{MJarrell96}
at the last iteration, we dramatically increased the number of sweeps
such that the number of bins, each containing 100 measurements,
is larger than $2L$.


\section{Results}
\label{SEC:Spectrum}

\begin{figure}[t]
\includegraphics[width=8cm, angle=0]{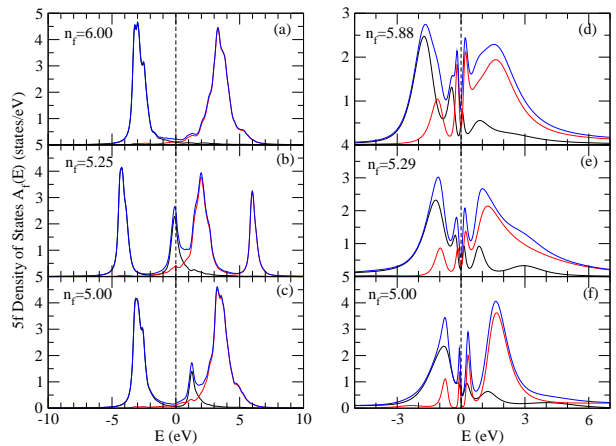}
\caption{(Color) The $5f$ spectral density calculated from
the DMFT(HI) [left panels] and DMFT(QMC) [right panels]  for
$\delta$-Pu for varying $5f$ site energy at $T=632$ K.
In all panels, the spectral density contributed from the $j=5/2$
and $j=7/2$ sub-shells are represented by the black and red lines,
respectively. The total $5f$ spectral density is represented by the
blue line. The energy $E=0$ denotes the position of the Fermi level.}
\label{FIG:nfDep}
\end{figure}


In Fig.~\ref{FIG:nfDep} we show the spectral density of
$\delta$-Pu at $T=632\,$K for various values of the $5f$
occupancy $n_f$.
The left three panels display underlying atomic properties
obtained from the DMFT(HI) calculations, consistent with a ground
state $\alpha |f_{5/2}^5\rangle + \beta |f_{5/2}^6\rangle$, where
$\vert\alpha\vert^2\!+\!\vert\beta\vert^2=1$ and $n_f = 5\!+\!\vert
\beta\vert^2$.  For $n_f\!=\!6$ in Fig.~\ref{FIG:nfDep}(a), one gets
a single lower ($f_{5/2}^6 \!\rightarrow\! f_{5/2}^5$) and upper
($f_{5/2}^6 \!\rightarrow\! f_{5/2}^6 f_{7/2}^1$) Hubbard band.
For $n_f\!=\!5$ in Fig.~\ref{FIG:nfDep}(c), however, one may still
add an electron to the $j\!=\!5/2$ subshell, so that in addition
to the lower ($f_{5/2}^5\! \rightarrow\!  f_{5/2}^4$) one has two,
spin-orbit split upper Hubbard bands ($f_{5/2}^5 \!\rightarrow\!
f_{5/2}^6$ and $f_{5/2}^5 \!\rightarrow\! f_{5/2}^5 f_{7/2}^1$).
For the mixed valent, nonintegral $n_f$ in Fig.~\ref{FIG:nfDep}(b),
the system behaves as an ensemble of $|\alpha|^2$ $f_{5/2}^5$
and $|\beta\vert^2$ $f_{5/2}^6$ configurations.  In the absence of
hybridization, this leads to four Hubbard bands in relative position
$4U_f$, $5U_f$, $5U_f\!+\!\Delta_f$, and $6U_f\!+\!\Delta_f$,
where $\Delta_f$ is the spin-orbit splitting, with areas of
$5\vert \alpha\vert^2$, $6\vert\beta\vert^2+\vert\alpha\vert^2$,
$8\vert\alpha\vert^2$, and $8\vert\beta\vert^2$, respectively.
The Fermi level (energy zero) should lie within the second Hubbard
band at $5U_f$ splitting the occupied ($6\vert\beta\vert^2$) and
empty ($\vert\alpha\vert^2$) parts, where the latter is the small
tail of the $j\!=\!5/2$ spectra (black) extending above the Fermi
level in Fig.~\ref{FIG:nfDep}(b).

For the more rigorous DMFT(QMC) results one sees, in the three
right-hand panels of Fig.~\ref{FIG:nfDep}, the expected transfer of
spectral weight away from the Hubbard bands into the quasiparticle
peak  at the Fermi level, which DMFT(HI) is incapable of describing.
However, in addition there also appears to be a shift of the outlying
DMFT(HI) Hubbard structure closer towards the Fermi level, while
at the same time the one Hubbard band already overlapping the Fermi
level [second from left in Fig.~\ref{FIG:nfDep}(b)] spreads away from
the Fermi level.  The overall affect is to give a more smooth and
systematic evolution of the DMFT(QMC) spectra with growing $n_f$ than
seen in DMFT(HI).  This is confirmed by an independent evaluation
of the DMFT(QMC) state density at the Fermi level via $A_j(0)
= (2j\!+\!1)(\beta/\pi)G_j(\tau\!=\!\beta/2)$.\cite{Trivedi99}
$A_{5/2}(0)$ smoothly decreases while $A_{7/2}(0)$ (and the total)
smoothly increase for increasing $n_f$ over the range $5\leq n_f
\leq 6$, consistent with the Fermi level passing more into the
$j\!=\!7/2$ part of a spin-orbit split  peak.  
Indeed, the MaxEnt results in Figs.~\ref{FIG:nfDep}(d--f) exhibit a 
splitting in the quasiparticle peak. On the one hand, 
  we suggest that it is due to the
spin-orbit interaction and the associated $j$-dependence
induced into the self-energy.  Such splitting appears in
the spectra for each $j$ as the two channels couple via
$\hat{H}^{0}(\mathbf{k})$.  A reanalysis using MaxEnt of earlier
DMFT(QMC) calculations~\cite{AKMcMahan05} for Ce and Pr exhibits
similar behavior, and since this is then independent of filling,
hybridization dips or gaps can be ruled out.  Such spin-orbit induced 
splitting
has been experimentally observed in photoemission experiments for
Ce,~\cite{Patthey85} although the Hund's rules exchange omitted in
Fig.~\ref{FIG:nfDep} (and discussed shortly) will have larger impact
for mutli-$f$ electron atoms.  
On the other hand,
we do not exclude the possibility that the splitting may be simply an artifact
of the analytical continuation technique used to obtain the spectral function
at the real energy axis.
Note that we find the quasiparticle
peak and its associated fine structure to both disappear for 
$\delta$-Pu at higher temperature, $1580\,$K, consistent with the 
$800\,$K Kondo energy that has been suggested.~\cite{JHShim06}


\begin{figure}[t]
 \includegraphics[width=8cm, angle=0]{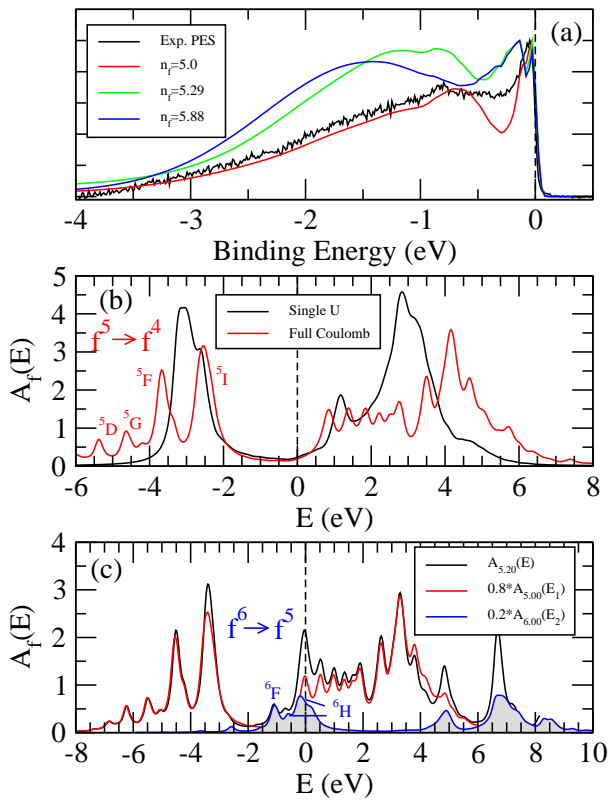}
\caption{(Color)  (a) Experimental  and theoretical photoemission spectrum.
(b) The $5f$ spectral density calculated from the
DMFT(HI) with only $F^{0}=U_{f}$ (black line) and full Coulomb (red line) interactions 
for $n_{f}=5.0$.  (c) The $5f$ spectral density calculated from the DMFT(HI) with 
full Coulomb interactions for $n_{f}=5.2$ and scaled/shifted ones 
for $n_{f}=5.0$ and 6.0. Here $E_{1}=E+0.87\;{\text{eV}}$ 
and $E_{2}=E-2.29\;{\text{eV}}$.
} \label{FIG:exp}
\end{figure}

A comparison of the DMFT(QMC) spectra with experimental photoemission
data~\cite{JMWills04,AJArko00} is shown in Fig.~\ref{FIG:exp}(a).  The $spd$
valence contribution was added to the calculated $5f$ spectra from
Figs.~\ref{FIG:nfDep}(d--f), and the total was broadened to reflect
the 60 meV instrument resolution and 15 K measurement temperature,
as well as the Lorentzian for the photohole lifetime determining the
natural linewidth, including quadratic scaling with binding energy.
For the experimental comparison, data from the 40.8 eV HeII-alpha
line were selected to obtain a photon energy range where
orbitals of interest have similar cross sections.~\cite{JMWills04} 
 While additional
work is certainly needed for an optimal comparison with the data,
Fig.~\ref{FIG:exp}(a) does reproduce the trend seen in the DMFT(QMC)
results of Figs.~\ref{FIG:nfDep}(d--f), where the lower Hubbard band
is seen to move to more negative energies with increasing $n_f$,
which lends experimental support for a value of
$n_f$ closer to 5.

The implications of the comparison in Fig.~\ref{FIG:exp}(a) do
not appear to be compromised by omitting Hund's rules exchange and
the associated term structure in
the theoretical results, where only the monopole Slater integral
$F^0\!=\!U_f$ has been incorporated.  This can be tested in
DMFT(HI), where the black curve in Fig.~\ref{FIG:exp}(b) is the
total $n_f\!=\!5.0$  $5f$ spectra of Fig.~\ref{FIG:nfDep}(c), while the
red curve now treats the full Coulomb interaction taking in addition
reasonable experimental values for $F^2$, $F^4$, and $F^6$.  In the
electron removal spectra one now sees the standard terms (labelled)
of the $f^4$ final state.  These agree in relative position and
intensity with earlier theoretical analysis,~\cite{FGerken83} and
serve only to broaden the lower Hubbard band to the negative-energy side,
leaving the positive-energy side of importance to the comparison in
Fig.~\ref{FIG:exp}(a) relatively unaffected.
While this DMFT(HI) test can not probe the impact of Hund's rules 
exchange on the quasiparticle peak near the Fermi level, one might 
anticipate a reduced effect after such fine structure has been 
appropriately broadened for comparison to photoemission data.

There has been much discussion of a three-peak structure
within about 1 eV of the Fermi level in the photoemission of
Pu systems: in thin layers of PuSe,~\cite{TGouder00} thin Pu
layers on Mg,~\cite{TGouder01,LHavela02} single 
crystal PuTe,~\cite{TDurakiewicz04} and suggested in thin 
film of PuSi$_{1.7}$~\cite{TGouder05} and 
PuN.~\cite{LHavela03}  The presence of the
structure in thin Pu layers (a few ML) was interpreted as a result
of localization effects due to low dimensionality overcoming
the itinerant character of 5f electrons.~\cite{TGouder01}
More recently it has been suggested that such structure may be
evident in $\delta$-Pu metal itself.~\cite{TGouder05,JHShim06} Such
structure may also be seen in DMFT(HI) calculations including
the full Coulomb interaction. In Fig.~\ref{FIG:exp}(c),
the black curve corresponds to a DMFT(HI) calculation with the
$5f$ site energy adjusted so $n_f\!=\!5.2$.  For comparison the
red (blue/shaded) curve shows the $n_f\!=\!5.0$ ($6.0$) result
scaled by $|\alpha|^2\!=\!0.8$ ($|\beta|^2\!=\!0.2$) and shifted
so its lowest unoccupied (highest occupied) state is at the Fermi
level.  The three peaks discussed are shown labelled: $^6H_{5/2}$
essentially at the Fermi level, and then $^6H_{7/2}$ and $^6F_{5/2}$
moving below.  Given the agreement between the
DMFT(HI) spectra $A_{5.2}$ for $n_f\!=\!5.2$ with the composite
$0.8A_{5.0}+0.2A_{6.0}$ (addition not shown), it is evident that
especially the left two peaks arise from $f^5$ term structure
in the final state of $5f^6\rightarrow 5f^5$ electron removal, and
that the strength of these peaks might provide a direct measure
of the $5f^6$ admixture $\beta$ into the $\delta$-Pu ground state,
and therefore of $n_f$ itself.

There is little doubt that such an analysis is reasonable for some
Pu compounds, where precise atomic multiplet calculations were
found in agreement with photoemission data in both line
positions and intensities.~\cite{TGouder00}  Such agreement 
includes cases (cubic PuTe and PuSe) where the three-peak structure
is experimentally observed below the Fermi level approximately $100\,$meV.
The situation provided by a metallic environment as in $\delta$-Pu is
substantially different, however, and so the localized multiplet interpretation 
is not at all clear.
An obvious concern about the results in Fig.~\ref{FIG:exp}(c), for
example, is that the self-energy used in DMFT(HI)
is too atomic-like; the differences between
the DMFT(HI) and DMFT(QMC) curves in Fig.~\ref{FIG:nfDep}
add to this concern.

The nature of the peak right at the Fermi level in $\delta$-Pu
is clearer.  The low-temperature electronic specific heat
in bulk $\delta$-Pu metal is 
$64\pm 3$ mJ K$^{-2}$mol$^{-1}$.~\cite{JCLashley03} 
This large, heavy-fermion-like
value indicates significant $5f$ state density at the Fermi level
which is consistent with the kind of Kondo-like physics that
DMFT(QMC) can yield, albeit possibly at temperatures below the
present work.  
Corroborating
evidence comes from the photon energy dependent and temperature
dependent photoemission experiments performed on clean 
$\delta$-Pu
metal surfaces,~\cite{JMWills04,JJJoyce07,JJJoyce98} which 
strongly suggest that the first peak
in the $\delta$-Pu arises from $5f$ electrons that are  hybridized
with the conduction electrons, which is again a typical quasiparticle
 feature usually found for materials with an enhanced
electron mass.

\section{Summary}
\label{SEC:Summary}
We report a systematic LDA-DMFT study of the spectral
properties of $\delta$-Pu with varying $5f$ occupancy.
The strong sensitivity of the spectral properties to the $5f$
occupancy and the inherent ambiguity in defining 
the associated
$5f$ orbitals suggests the need to use $5f$ occupancy
as a fitting parameter in addition to the Hubbard $U$ for DMFT theories
of Pu and other materials.
By comparing with photoemission data, we conclude
that an ``open shell'' $5f^{5}$ configuration gives better agreement
with experiment than the hypothesized ``closed shell'' $5f^{6}$ 
case.~\cite{ABShick05,AOShorikov05,LVPourovskii05,AShick06} 
The DMFT(HI)
approach, which is accurate in the localized, atomic limit, gives
significantly different results from the more rigorous DMFT(QMC)
method, which confirms the presence of a significant itinerant character
in $\delta$-Pu, especially at the Fermi energy.
The ``3-peak'' structure seen within about 1 eV of the Fermi level
in the photoemission spectra of some Pu compounds as well as metallic
thin films, but which remains controversial for the bulk $\delta$-Pu
metal, may provide a measure of the admixture of $5f^6$
character in the ground state of these materials, and therefore an
additional indication of $n_f$.  However, it is not clear from the
present work whether this structure should wash out in a correlated
calculation for $\delta$-Pu, which rigorously treats both extended and intraatomic
exchange effects.

\acknowledgments 
We are grateful to  G. Kotliar, V. Oudovenko,
S. Y. Savrasov, K. Held, and C. D. Batista  for helpful
discussions, and to J. E. Gubernatis and M. Jarrell for making
available their maximum entropy code. We acknowledge the support
of the US DOE at LANL under Contract No. DE-AC52-06NA25396, and 
at LLNL under Contract No. W-7405-Eng-48.


\begin{thebibliography}{99}
\bibitem{ASAresults} In DMFT(QMC) calculations using our constrained
occupation values of both $\varepsilon_f$ and $U_f$ along with an
LMTO-ASA $\hat{H}^{\text{LDA}}(\mathbf{k})$, we also find $n_f \sim
5$, in agreement with the spectroscopic-based conclusions of the
present paper.
\end{thebibliography}
\end{document}